\documentstyle[floats,aps,preprint]{revtex}
\begin{document}
\draft
\preprint{\vbox{{\hbox{SOGANG-HEP 266/99}
                 \hbox{hep-th/9911085}  }}}
\title{Noncommutative open strings from Dirac quantization}
\author{Won Tae Kim\footnote{electronic address:wtkim@ccs.sogang.ac.kr}
      and John J. Oh\footnote{electronic address:john5@gravity.sogang.ac.kr}}
\address{Department of Physics and Basic Science Research Institute,\\
         Sogang University, C.P.O. Box 1142, Seoul 100-611, Korea}
\date{\today}
\maketitle
\bigskip 
\begin{abstract}
We study Dirac commutators of canonical variables
on D-branes with a constant Neveu-Schwarz 2-form field
by using the Dirac constraint quantization method, 
and point out some subtleties appearing in previous 
works in analyzing constraint structure of the brane system. 
Overcoming some ad hoc procedures, we obtain 
desirable noncommutative coordinates exactly 
compatible with the result of the conformal field theory 
in recent literatures. Furthermore, we find interesting
commutator relations of other canonical variables.  
\end{abstract}

\bigskip

\newpage

Noncommutative geometry arises in D-branes on
constant antisymmetric tensor fields \cite{cds},
and gauge theories on the noncommutative space 
have been studied by Seiberg and Witten \cite{wit,sw}.
The fundamental commutator relations of open string coordinates $x^i$ on the branes are 
represented by noncommutativity, 
$[x^i(0,\tau),x^j(0,\tau)]=i \theta^{ij}$ where
$\theta^{ij}$ is an antisymmetric constant
tensor depending on the background constant
fields. Various aspects of the noncommutativity on the branes 
are extensively studied in Refs. \cite{dh,jab,aas2,aj,ch2}.  
Furthermore, 
the noncommutativity in the Matrix Model 
with non-trivial three form field has been 
observed in Ref.\cite{cds}.
Especially, the noncommutativity 
between D-brane coordinates is shown 
in terms of Green's function method \cite{sw}
and string mode expansion method \cite{aj,ch2}.

In recent studies \cite{aas,ss,ch}, 
it has been suggested that noncommutative 
coordinates on the brane coupled to constant antisymmetric 
background fields can be derived from the Hamiltonian 
formulation of the system by treating the mixed boundary 
condition as a primary constraint. 
Another intriguing  point
adopted in Refs. \cite{aas,ss} for the Hamiltonian formulation
is to discretize the string world
sheet as $X_n$ $(n=0,1,2 \cdots)$
where $X_0$ describes the boundary of the open string and 
the others are for the bulk, and the noncommutative structure of 
the boundary and the bulk part can be studied 
respectively.     
On the other hand, the stability of the
primary constraint with respect to time gives infinite
number of secondary constraints 
where the Lagrangian multiplier $u^i$ implemented by the primary constraint 
was determined
as $u^i=0$. 
The resulting constraints of the primary
and the secondary constraints form second class algebra,
and Dirac brackets between the coordinates on the brane
yield the noncommutative coordinates.
At first sight, however, 
the noncommutativity seems to appear both
at the boundary and partially at the bulk\cite{aas,ss}. 
In Ref. \cite{ch}, the Dirac quantization of this system
was performed with the infinite number of secondary constraints
and obtained the noncommutative coordinate relations at
the boundary, however, it is more or less formal and the regularization of $\delta$-function is needed.
In Refs. \cite{aas,ss,ch}, the essential subtlety is due to the fact that the canonical Hamiltonian 
instead of the primary 
Hamiltonian has been used 
in the stability condition of the primary constraint, 
which it is different from the conventional Dirac quantization 
procedure \cite{dir}.

In this paper, we shall reconsider 
Dirac constraint quantization
of the brane on the constant antisymmetric backgrounds and obtain
consistent noncommutative commutators of canonical
variables including momenta. Compared to the previous works \cite{aas,ss,ch},
the crucial difference comes from the choice of the nonvanishing 
multiplier in the primary Hamiltonian, which is determined by
the stability condition of the primary constraint.   
In our case, there are no more secondary constraints,
instead, the single primary constraint itself forms
the second class constraint algebra, which gives
various kinds of interesting commutators as well as
the expected noncommutative coordinates. 
In our analysis,
the coordinates in the bulk part are definitely commutative, 
which is different from some earlier works \cite{aas,ss}.   
  
Before proceeding, 
let us exhibit self-dual constraints appearing in 
chiral quantum mechanics \cite{bs} as a simple
illustration of noncommutative coordinates, which is very 
similar to the constraint system in the
open string theory.
Then the Lagrangian is given by
\begin{equation}
  \label{lagrangian}
  L = - \frac{e}{2c} B \dot{x}^{i} \epsilon_{ij} x^{j} +e \phi(x),
\end{equation}
where it is obtained from a charged point particle coupled 
to the static potential $\phi$ on
a strong magnetic field background in three dimensions.
The canonical momenta are
\begin{equation}
  \label{canomom}
  p_{i} = \frac{\partial L}{\partial \dot{x}^{i}} = 
-\frac{e}{2c} B \epsilon_{ij} x^{j},
\end{equation}
which becomes a primary constraint in this theory.
This is called simply self-dual constraint in that
the momenta are represented by the coordinates
\begin{equation}
  \label{primcon}
  \omega_{i} = p_{i} + \frac{e}{2c} B\epsilon_{ij} x^{j} \approx 0.
\end{equation}
Considering a primary Hamiltonian \cite{dir},
\begin{equation}
  \label{primham}
  H_{p} = H_{c} + u^{i} \omega_{i}, 
\end{equation}
where $H_{c} = -e \phi$ and $u^{i}$ is a multiplier,
the stability condition of the primary constraint
with respect to time yields
\begin{eqnarray}
  \label{lagmul}
  \{\omega_{i} , H_{p}\} &=& e \partial_i \phi +
              \frac{e}{c} B  \epsilon_{ij} u^j, \nonumber \\
         &\approx & 0.
\end{eqnarray}
If the magnetic field is nonvanishing, it is possible 
to fix the multiplier, whereas for $B \rightarrow 0$ the
additional constraint called a secondary constraint can
be generated, so the primary constraint 
together with the secondary constraint
form second class constraint algebra. However, in our case, the
primary constraint itself becomes a second class
constraint and the multiplier can be naturally fixed.

Now the consistent brackets with the primary
constraint (\ref{primcon}) are defined as
\begin{equation}
[{\cal A}, {\cal B}]=i\{{\cal A},{\cal B}\}
+i\{{\cal A},\omega_i\}\frac{c}{eB}\epsilon^{ij}
     \{\omega_j, {\cal B}\},
\end{equation}
where ${\cal A}$ and ${\cal B}$ are canonical variables
and the resulting commutators are
\begin{eqnarray}
  \label{dirbracket}
  & &[x^{i}, x^{j}] = - i\frac{c}{eB} \epsilon^{ij}, \nonumber \\
  & &[x^{i}, p^{j}] = \frac{i}{2}g^{ij},\nonumber \\
  & &[p^{i}, p^{j}] = - i\frac{e}{4c} B \epsilon^{ij}.
\end{eqnarray}
The commutators between the coordinates are noncommuting 
as well as those of the momenta. 
In fact, this
curious feature is due to the constraint (\ref{primcon}) to be
imposed on the phase space, which will be
studied in the context of the open string
theory. 

We now consider the open string theory coupled to 
constant Neveu-Schwarz 2-form fields
and $U(1)$ gauge field background given by
\begin{eqnarray}
  \label{action}
  S&= &\frac{1}{4\pi \alpha'} \int d{\sigma}^2 \left[\partial_{a}X^{i}\partial^{a}X_{i} + 2\pi \alpha'  B_{ij} \epsilon^{ab} \partial_{a}X^{i} 
\partial_{b} X^{j} \right] \nonumber \\
&&+ \int d \tau A_i \partial_{\tau} X^{i}|_{\pi}
- \int d \tau A_i \partial_{\tau} X^{i}|_{0}
\end{eqnarray}
where it has a local $U(1)$ gauge invariance. 
If both ends of a string attached to the same brane, the last
two boundary term in Eq.(\ref{action}) can be written as
$ -\frac{1}{2\pi \alpha'} \int d \tau 
F_{ij} \epsilon^{ab} \partial_{a}X^{i} 
\partial_{b} X^{j}$, and the action (\ref{action})
becomes
\begin{equation}
S= \frac{1}{4\pi \alpha'} \int d{\sigma}^2 \left[\partial_{a}X^{i}\partial^{a}X_{i} + 2\pi \alpha'  {\cal F}_{ij} \epsilon^{ab} \partial_{a}X^{i} 
\partial_{b} X^{j} \right]
\end{equation}
and the two form field ${\cal F} = B - F = B - dA$ is 
invariant under both $U(1)$ and $\Lambda$ transformation 
defined as $B \rightarrow B+ d \Lambda$ and $A \rightarrow A +\Lambda$. 
We simply take $F_{ij}=0$ in Eq. (\ref{action}) for convenience.
 
Varying the action (\ref{action}) gives the equation of motion, 
$ \partial_{a}\partial^{a} X^{i} = 0$ with the mixed boundary
condition 
\begin{equation}
g_{ij}\partial_{\sigma}X^{j} + 2\pi \alpha' B_{ij} \partial_{\tau}X^{j} |_{\sigma = 0,\pi} = 0. \label{bc}
\end{equation}
Without the constant background fields $B_{ij}$, 
the boundary conditions
(\ref{bc}) are reduced to Neumann boundary conditions 
of the open string theory. 
To distinguish the boundary and the bulk, 
we discretize our action (\ref{action}) 
along the $\sigma$ parameter\cite{aas}
and the resulting discretized Lagrangian becomes
\begin{equation}
  \label{dislagrang}
  L = \frac{1}{4 \pi \alpha'} \sum_{n}\left[ \epsilon(\dot{X}^{i}_{n})^2 - 
\frac{1}{\epsilon} \left(X^{i}_{n+1} - X^{i}_{n}\right)^2 + 
4\pi \alpha' B_{ij} \dot{X}^{i}_{n} \left(X^{j}_{n+1} - X^{j}_{n}\right)\right]
\end{equation}
where we take equal spacing, $\int d \sigma =\epsilon \sum_{n}$
where the overdot means the derivative with respect to the time-like
parameter $\tau$.
And we also discretize the mixed boundary condition (\ref{bc}) as
\begin{eqnarray}
  \label{dismixbc}
  \frac{g_{ij}}{\epsilon}\left(X_{1}^{j} - X_{0}^{j}\right) + 2\pi\alpha'B_{ij} \dot{X}_{0}^{j} = 0
\end{eqnarray}
where it is simply denoted at the one boundary $\sigma=0$ for
convenience. 
>From the discretized Lagrangian (\ref{dislagrang}), 
we obtain canonical momenta given as
\begin{equation}
  \label{mom}
  2\pi \alpha' P_{ni} = \left[ \epsilon \dot{X}_{ni} + 2\pi \alpha'B_{ij} (X_{n+1}^{j} - X_{n}^{j})\right],
\end{equation}
where $n=0$ and $n=1,2,\cdots$ denote the coordinates 
on the brane and the string bulk,
respectively. 
According to the usual Dirac's procedure \cite{dir}, 
one can define the mixed boundary condition (\ref{dismixbc})
as a primary constraint by using Eqs.(\ref{dismixbc}) and (\ref{mom}), 
\begin{equation}
  \label{primarycon}
  \Omega_{i} = \frac{1}{\epsilon}\left[(2\pi\alpha')^2B_{ij} 
P_{0}^{j} -  (2\pi\alpha')^2B_{ij}B^{jk}(X_{1k}-X_{0k}) 
+ g_{ij}(X_{1}^{j} -X_{0}^{j})\right] \approx 0. 
\end{equation}
Then the primary hamiltonian 
can be constructed by performing the Legendre transformation 
of Eq.(\ref{dislagrang})
and by introducing the primary constraint implemented by the
multiplier $u^{i}(\tau)$, 
which is given as
\begin{eqnarray}
  \label{primham}
  H_{p} &=& H_{c} + u^{i} \Omega_{i} \nonumber \\
        &=& \frac{1}{4\pi\alpha'\epsilon} \sum_{n}\left[ (2\pi\alpha')^2 \left(P_{n}^{i} - B^{ij}(X^{n+1}_{j} - X^{n}_{j})\right)^2 
+(X_{n+1}^{i}-X_{n}^{i})^2\right] + u^{i} \Omega_{i},
\end{eqnarray}
where $H_{c}$ is a canonical Hamiltonian.

Using Poisson brackets defined by
\begin{eqnarray}
  \label{poisson}
  \{X_{n}^{i}, X_{m}^{j}\} &=& 0 = \{P_{n}^{i}, P_{m}^{j}\},\nonumber \\
  \{X_{n}^{i}, P_{m}^{j}\} &=&  g^{ij}\delta_{nm},
\end{eqnarray} 
the time evolution of the primary constraint 
(\ref{primarycon}) yields 
\begin{eqnarray}
  \label{multi}
  \{ \Omega_{i}, H_{p}\} = &-&\frac{2\pi\alpha'}
{\epsilon^2}\left[B_{ij}\left((2\pi\alpha')^2P_{0}B - 
(2\pi\alpha')^2B^2 (X_{1}-X_{0}) - (X_{1}-X_{0})\right)^{j}\right.\nonumber \\
&-&\left(g-(2\pi\alpha')^2B^2\right)_{ij}\left(P_{0} -B(X_{1}-X_{0})\right)^{j}\nonumber \\
&+&\left(g-(2\pi\alpha')^2B^2\right)_{ij}\left(P_{1} -B(X_{2}-X_{1})\right)^{j}\nonumber \\
&+& \left. 2(2\pi\alpha')B_{ij}(g-(2\pi\alpha')^2B^2)^{j}_{k}u^{k}(\tau)\right] 
\end{eqnarray}
where this condition determines 
the multiplier rather than 
it generates additional constraints since
$B_{ij}$ is invertible\cite{sw}.
In Refs.\cite{aas,ss,ch}, however, infinite chain of constraints 
appeared as secondary constraints
and it is different 
from the usual Dirac quantization procedure.
The essential reason why the multiplier $u^{i}$
can be fixed is due to the Poisson algebra of the primary constraint
(\ref{primarycon}) which yields second class constraint 
algebra as
\begin{equation}
 \{\Omega_{i}, \Omega_{j}\}=\frac{2}{\epsilon^2}(2\pi\alpha')^2
\left[(g-2\pi\alpha'B)B(g+2\pi\alpha'B)\right]_{ij}.
\end{equation}
It would be interesting to note that
for $B_{ij}\rightarrow 0$ the primary constraint 
becomes first class constraint, in that case,
the time evolution of the primary constraint by
using the primary Hamiltonian gives secondary 
constraint. These constraints form the second class constraint
algebra and the Lagrangian multiplier can 
be fixed.    
  
Returning to our analysis, let us now construct the Dirac matrix defined by
\begin{equation}
  \label{diracmatrix}
  C_{ij} = \{\Omega_{i}, \Omega_{j}\},
\end{equation}
and the inverse matrix can be obtained as
\begin{equation}
\label{diracmatrix}
   {C^{ij}}^{-1} = \frac{\epsilon^2}{2(2\pi\alpha')^2} \left[\frac{1}{(g+(2\pi\alpha')B)}\frac{1}B\frac{1}{(g-(2\pi\alpha')B)}\right]^{ij}.
\end{equation}
Then we can calculate 
Dirac commutators straightforwardly by means of the following definition 
\begin{equation}
  \label{diracbracket}
  [A, B] = i\{A,B\}-i\{A, \Omega_{i}\}{C^{ij}}^{-1}\{\Omega_{j}, B\}.
\end{equation}
The small constant parameter $\epsilon$ can be canceled out
in Dirac commutators and commutator relations
are valid for the prescription $\epsilon \rightarrow 0$
after finishing all Dirac procedures.
The resulting commutators at the boundary ($\sigma=0$) of the
open string are
\begin{eqnarray}
 & & [X_{0}^{i}, X_{0}^{j}] = -\frac{i}{2}(2\pi\alpha')^2\left[\frac{1}{(g+(2\pi\alpha')B)}B\frac{1}{(g-(2\pi\alpha')B)}\right]^{ij},
\label{x0x0}\\
 & & [X_{0}^{i}, P_{0}^{j}] =\frac{i}{2}g^{ij},\label{x0p0} \\
& & [P_{0}^{i}, P_{0}^{j}] = \frac{i}{2(2\pi\alpha')^2}\left[(g+(2\pi\alpha')B)\frac{1}B(g-(2\pi\alpha')B)\right]^{ij}. \label{p0p0}
\end{eqnarray}
If we redefine the phase space variables as
$ X_{0}^{i} \rightarrow \frac{1}{\sqrt{2}}X_{0}^{i}$, then
the Dirac commutator of the coordinates (\ref{x0x0}) is equivalent to the
result derived from the propagator by Seiberg and Witten 
in Ref. \cite{sw}. The commutator relations between the
canonical momenta are nontrivial, which is similar to the
form of the point particle case on the presence of
the magnetic field. 
Note that for a point particle limit of $\alpha'\rightarrow 0$, 
the commutation relation for the coordinates (\ref{x0x0})
is reduced to $[x^{i}(\tau), x^{j}(\tau)] = i(B^{-1})^{ij}$ 
whose form is the same with that of Eq.(\ref{dirbracket}) with a constant redefinition.

On the other hand, at the nearest point from the boundary for $n=1$, 
the Dirac commutators are
\begin{eqnarray}
& & [X_{1}^{i}, X_{1}^{j}] = 0, \label{x1x1} \\
& & [X_{1}^{i}, P_{1}^{j}] = ig^{ij}, \label{x1p1} \\
& & [P_{1}^{i}, P_{1}^{j}] = \frac{i}{2(2\pi\alpha')^2}\left[(g+(2\pi\alpha')B)\frac{1}B(g-(2\pi\alpha')B)\right]^{ij}.\label{p1p1}
\end{eqnarray}
Note that commutator relations between the coordinates $X_1^i$
in Eq.(\ref{x1x1}) imply that they are commuting each other,
which is in contrasted with the result of 
the noncommutative coordinates \cite{aas}. 
Next, the cross commutation relations are also given as
\begin{eqnarray}
\label{cross}
& & [X_{0}^{i},X_{1}^{j}]  = 0, \nonumber \\
& & [X_{0}^{i}, P_{1}^{j}] =  \frac{i}{2}g^{ij},\nonumber \\
& & [X_{1}^{i}, P_{0}^{j}] = 0,\nonumber \\
& & [P_{0}^{i}, P_{1}^{j}] = -[P_{0}^{i}, P_{0}^{j}].
\end{eqnarray}
Furthermore, the other 
Dirac commutators for the string bulk, i.e. $n=2,3,\cdots$, 
are the same with the usual Poisson brackets, 
\begin{eqnarray}
  \label{bulkcomm}
   & &[X_{n}^{i}, X_{m}^{j}] = 0 = [P_{n}^{i}, P_{m}^{j}],\nonumber \\
   & &[X_{n}^{i}, P_{m}^{j}] = i\delta_{nm} g^{ij}.
\end{eqnarray}
>From Eqs.(\ref{x0x0})-(\ref{bulkcomm}),
we find that the boundary coordinate $X_0^i$ is nontrivially correlated
with the other canonical variables for $n=0,1$,
whereas the coordinate $X_1^i$ is commuting with other
canonical variables except its conjugate momentum $P_1^i$.

The similar commutators for the $\sigma=\pi$ boundary of the D-brane 
are given as
\begin{eqnarray}
  \label{diracbrackets}
 & & [X_{\pi}^{i}, X_{\pi}^{j}] = \frac{i}{2}(2\pi\alpha')^2\left[\frac{1}{(g+(2\pi\alpha')B)}B\frac{1}{(g-(2\pi\alpha')B)}\right]^{ij},\nonumber \\
 & & [X_{\pi}^{i}, P_{\pi}^{j}] =\frac{i}{2}g^{ij},\nonumber \\
 & & [X_{\pi-\epsilon}^{i}, X_{\pi-\epsilon}^{j}] = 0,\nonumber \\
 & & [P_{\pi}^{i}, P_{\pi}^{j}] = -\frac{i}{2(2\pi\alpha')^2}\left[(g+(2\pi\alpha')B)\frac{1}B(g-(2\pi\alpha')B)\right]^{ij}\nonumber \\
 & & \qquad \ \ \ \ =[P_{\pi-\epsilon}^{i}, P_{\pi-\epsilon}^{j}]\nonumber\\
 & & \qquad\ \ \ \ =-[P_{\pi}^{i}, P_{\pi-\epsilon}^{j}],\nonumber\\
 & & [X_{\pi}^{i}, P_{\pi-\epsilon}^{j}] = \frac{i}{2}g^{ij},\nonumber \\
 & & [X_{\pi-\epsilon}^{i}, P_{\pi-\epsilon}^{j}] = ig^{ij},\nonumber \\
 & & [X_{\pi-\epsilon}^{i}, P_{\pi}^{j}] = 0 = [X_{\pi}^{i},X_{\pi-\epsilon}^{j}].
\end{eqnarray}
As a result, the noncommutativity of the coordinates 
emerges only on the branes not in the string bulk. 

On the other hand, commutation relation of 
the center of mass(CM) coordinate 
is calculated as 
\begin{eqnarray}
  \label{cmcommut}
  [X_{\rm CM}^{i},X_{\rm CM}^{j}] &=& \left
(\frac{\epsilon}{\pi}\right)^2 \sum_{n}\sum_{m}
[X_{n}^{i}, X_{m}^{j}]\nonumber \\
&=& \left(\frac{\epsilon}{\pi}\right)^2 
\left([X_{0}^{i},X_{0}^{j}]+[X_{\pi}^{i},X_{\pi}^{j}]\right)
\nonumber\\
&=&i\left(\frac{\epsilon}{\pi}\right)^2
 (2\pi\alpha')^2  \left[((\tilde{g}-(2\pi\alpha')^2{\tilde{B}}^2)^{-1}\tilde{B})^{ij} -((g-(2\pi\alpha')^2B^2)^{-1}B)^{ij}\right]
\end{eqnarray}
where we used the definition of
the center of mass coordinate as
$  X_{\rm CM}^{i}=\frac{1}{\pi}\int_{0}^{\pi} d\sigma X^{i}
 =\frac{\epsilon}{\pi} \sum_{n} X_{n}^{i}$.
The only boundary commutators 
at $\sigma=0$ and $\sigma=\pi$ contribute 
to the center of mass commutation relation.
The tilde fields 
are defined at the $\sigma=\pi$ boundary of the brane.
As discussed in \cite{aas}, 
for the $MM$ brane system, 
the center of mass commutator 
is vanishing while for the $MM'$ brane system 
it is vanishing in this discretized version. In Ref. \cite{aas},  
noncommutative coordinate relations of the boundary and the bulk 
contribute to the calculation of the commutation relation of
the center of mass coordinates.
In our case, if $\epsilon \rightarrow 0$, then the
commutator relation of the center of mass coordinates (\ref{cmcommut})
is always vanishing. This fact seems to be
plausible in that the CM coordinate belongs the bulk part 
which is commuting.  

It seems to be appropriate to comment on 
the nontrivial commutator relations between
coordinates and momenta. At first sight, 
one might wonder why
the coordinate $X_0^i$ is related to the momentum 
$P_1^i$ through the commutation relation as seen in Eq.(\ref{cross}).
As a matter of fact, this is due to the
structure of the primary constraint (\ref{primarycon}) 
since it contains the canonical variables $X_1^i$ and
affects the Dirac bracket. 
So the constraint is not only defined at the boundary point $X_0^i$
but also it is related to the next point $X_1^i$ through
the derivative at $\sigma=0$ in the mixed boundary condition (\ref{primarycon}). Strictly speaking, the coordinates are
noncommutative only at the boundary, however, the
canonical variables including momenta at the boundary $\sigma=0$
are correlated 
up to the next canonical variables.  
As for the canonical momenta, they are noncommutative
for $n=0,1$, whereas for $n=2,3, \cdots$ they are commuting.
  
In conclusions, 
we have shown that 
the noncommutative coordinates can be obtained at
the boundary of open strings on
the constant antisymmetric fields, while the
bulk coordinates as expected follow the 
usual commuting relations, which has been
studied in the context of
the Dirac constraint quantization method
compatible with the propagator method
used in the conformal field theory. 
\vspace{3cm}
 
{\bf Acknowledgments}\\
We would like to thank M.M. Sheikh-Jabbari for exciting
discussions and P.-M. Ho for careful reading our manuscript and
helpful suggetions, and  M.S. Yoon for helpful comments.
This work was supported by Brain Korea 21 program of 
Korea Research Foundation(1999).


\end{document}